\documentclass[aps,twocolumn,superscriptaddress,floatfix,showpacs]{revtex4}
\pdfoutput=1
\usepackage[latin1]{inputenc}
\usepackage{graphicx} \usepackage{amsmath, bbm,amssymb, color,wasysym}
 
\begin{document}

\bibliographystyle{prsty}

\title{Quantum nondemolition-like, fast measurement scheme for a superconducting qubit}

\author{I. Serban}  

\affiliation{Physics Department, Arnold Sommerfeld Center for
Theoretical Physics, and Center for NanoScience, \\
Ludwig-Maximilians-Universit\"at, Theresienstrasse 37, 80333 Munich,
Germany} 

\affiliation{IQC and Dept.~of Physics and Astronomy, University of
Waterloo, 200 University Ave W, Waterloo, ON, N2L 3G1, Canada}  

\author{B.L.T. Plourde}
\affiliation{Department of Physics, Syracuse University, Syracuse, NY 13244-1130}

\author{F.K. Wilhelm}  
\affiliation{IQC and Dept.~of Physics and Astronomy, University of Waterloo, 200 University Ave W, Waterloo, ON, N2L 3G1, Canada}  

\date{ \today}

\begin{abstract} 
We present a measurement protocol for a flux qubit coupled to a dc-Superconducting QUantum Interference Device (SQUID),  representative of any two-state system with a controllable coupling to an harmonic oscillator quadrature, which consists of two steps. First, the qubit state is imprinted onto the SQUID via a very short and strong interaction. We show that at the end of this step the qubit  dephases completely, although the perturbation of the measured qubit observable during this step is weak. In the second step, information about the qubit is extracted by measuring the SQUID. This step can have arbitrarily long duration, since it no longer induces qubit errors.
\end{abstract}


\maketitle
\section{Introduction} 

The quantum measurement postulate is often viewed as the most intriguing assumption of quantum physics. Much of it has been demystified by the study of the physics of quantum measurements. The dynamics of the measurement process can be described by a coupled many-body Hamiltonian, consisting  of the system to be measured and the detector with a heat bath component \cite{Braginsky95,Caves80}.  Thus, the measurement process can be investigated using the established tools of quantum mechanics of open systems \cite{Nato06II,Weiss99, Keil01,Lindblad76}.

Most interest has been focused on the physics of weak measurements, where the system-observer coupling can be treated within perturbation theory. Famously, this research has shown that only a certain class of measurements satisfy von Neumann's quantum measurement postulate \cite{Cohen92, Neumann55} and indeed project the system wavefunction onto an eigenstate of the measured observable. Measurements of this type are termed quantum nondemolition (QND) measurements. Within the weak measurement paradigm, the QND regime is achieved when the measured observable is a constant of the free motion and commutes with the system-detector coupling Hamiltonian. Weak QND measurements have been investigated in various systems, ranging from spins to oscillators and even photons \cite{Bulaevskii04,Averin02,Jordan05,Jordan06,Milburn83,Sanders89,Brune90,Boulant07,boissonneault08}.

The dynamics of the weak measurement process has practical relevance in the context of quantum computing. Specifically, superconducting qubits have been proposed as building blocks of a scalable quantum computer \cite{Makhlin01,Devoret04,Nato06I,Clarke08}, and a fast measurement with a high resolution and visibility is important for readout and also for error correction.

There are a variety of different measurement techniques used in superconducting qubits.  
Weak measurements can be performed using single-electron transistors \cite{Makhlin01}. A different approach is the switching measurement, where the detector switches out of a metastable state depending on the state of  the qubit \cite{Science00,EPJB03,Vion02,Martinis02,steffen:050502}.  Such switching measurements have been a quite successful readout scheme for many superconducting qubit experiments to date. However, the dissipative nature of the switching process imposes limitations on the measurement speed and perturbs the qubit state.

A QND measurement could be achieved by using a pointer system, and measuring one of its observables influenced by the state of the qubit \cite{PRBR033}. Recent developments of such detection schemes, using an oscillator as the pointer, have led to vast improvements \cite{Lupascu04, Lupascu07, Lee05, Blais04, Schuster05, Wallraff05, Metcalfe07} over previous measurement protocols. 

It has previously been shown \cite{Lougovski06,Bastin06,Santos07,Storcz06} that infinitesimally short interaction between a qubit and an oscillator is sufficient to imprint information about the state of the oscillator onto the qubit. The similar idea of using a short interaction to transfer information about the qubit into the oscillator has been used \cite{measurement07} in a dispersive readout scheme. In this case, after a short interaction, the state of the oscillator contains information about the qubit which can be extracted by further measuring one of its observables, for example, momentum. However, this scheme did not take possible bit flip errors into account. These errors may occur in the short yet finite time when the qubit is in contact with its environment. Thus, the full power of a quasi-instantaneous measurement has not yet been explored. 

In this paper we describe the effect of an ideally extremely short and arbitrarily strong interaction of a qubit with its environment (consisting of a weakly damped harmonic oscillator). We investigate the back-action on the qubit when the measured observable does not commute with the Hamiltonian describing the interaction with the environment, and study how close this result approximates the QND measurement. 

We study a setup consisting of a flux qubit inductively coupled to a dc-SQUID magnetometer. The flux qubit consists of a superconducting loop with three Josephson junctions \cite{Mooij99,Orlando99}. For flux bias near odd half-integer multiple of $h/2e$, the qubit is represented by two circulating current states with opposite directions. During the entire measurement process the SQUID is coupled to measurement circuitry, with associated dissipative elements.  However, it never switches out of the zero dc-voltage state. The qubit-SQUID interaction of arbitrary strength is turned on only for a short time by applying a very short bias current pulse to the latter. During this time, information about the qubit is imprinted onto the SQUID and can later be extracted from it during the post-interaction phase by monitoring voltage oscillations across the device. When the current pulse is switched off, the qubit-SQUID interaction ideally vanishes and the environment no longer perturbs the qubit. Thus, one can afford a long time to measure the SQUID and determine the state of the qubit. 

In section \ref{method}, following Ref.~\cite{Tian02}, we model the qubit-SQUID system by a two-level system linearly coupled to a dissipative oscillator.  We describe the evolution of this system by means of a
master equation in the Born-Markov approximation \cite{Blum96}, valid for the underdamped SQUID. In section \ref{results} we discuss the qubit-oscillator evolution during both interaction and post-interaction phases.  We study the qubit dephasing and relaxation during the interaction phase. We show that, at the end of this phase, the qubit appears completely dephased. In other words, the qubit has been measured and its information has been transferred in the form of a classical probability to the oscillator. During the same time interval, we find that qubit relaxation has remained negligible. For the post-interaction phase we describe the evolution of the oscillator under the
influence of the environment, starting from the state prepared by the interaction with the qubit. Technically, extracting the qubit information amounts to measuring the amplitude of the ringdown of the oscillator momentum. In section \ref{implementation} we discuss some of the details involved with implementing this measurement scheme.

\section{Model and method}\label{method} 
\begin{figure}[!h]
  \includegraphics[width=\columnwidth]{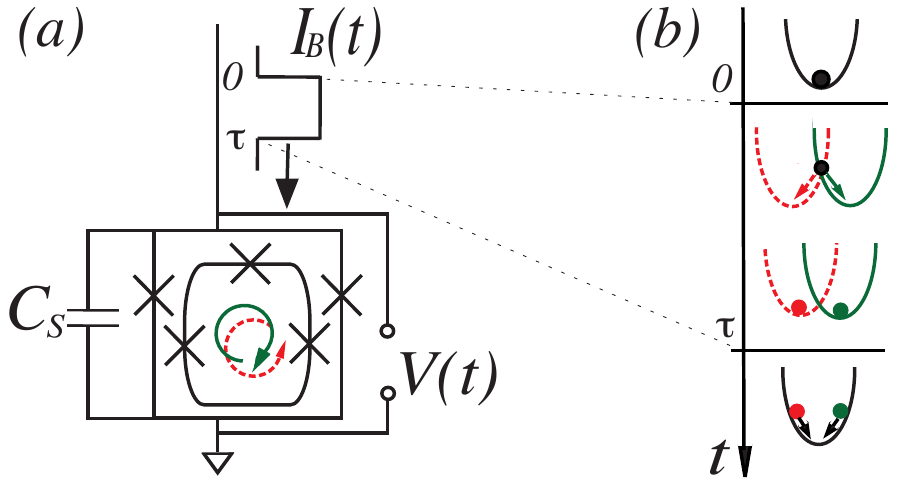}
\caption{(Color online). 
(a) Simplified circuit consisting of a flux qubit inductively coupled to a SQUID with two identical junctions and shunt capacitance $C_S$. The SQUID is driven by an bias step-like dc pulse $I_B(t)$ and the voltage drop $V(t)$ is measured by a device with internal resistance $R$. 
(b) Illustration of the measurement scheme: coupling ($t=0$) and decoupling ($t=\tau$) of the qubit and the SQUID (oscillator) and the  evolution of a point of mass in the transition of potential from one harmonic oscillator to a superposition of two displaced oscillators and back. The dashed (red) and the continuous (green) lines correspond to the different states of the qubit.}\label{circuit}
 \end{figure}

We study a flux qubit inductively coupled to a dc-SQUID, with one possible setup shown schematically in Fig.~\ref{circuit} (a). We describe a more detailed setup for implementing this scheme in section \ref{implementation}.

The SQUID is characterized by a two-dimensional washboard potential for the two independent phases corresponding to the two junctions \cite{Lefevre92}. Their sum couples to bias current driven through the SQUID, while the difference of phases couples to the magnetic flux applied to the SQUID. The small oscillations in these two directions can have vastly different characteristic frequencies. In particular, a small geometric inductance and a low critical current can make the flux mode frequency large while a shunt capacitor can lower the bias current mode frequency substantially. In the limit of very different frequencies, one can approximate the SQUID dynamics as that of a one-dimensional oscillator in the bias current direction, with the position of the oscillator minimum dependent on both $I_B$ and the total flux coupled to the SQUID which, for example, could vary depending on the state of the qubit.

The setup of Fig.~\ref{circuit} (a) can be described by the effective Hamiltonian \cite{Tian02}
\begin{eqnarray} 
\hat{H}&=&\hat{H}_S+\hat{H}_I+\hat{H}_B,\\
\hat{H}_S&=&\hbar w\hat{\sigma}_z+\hbar\delta\hat{\sigma}_x+\hbar\Omega(\hat{a}^\dagger
\hat{a}+1/2)\nonumber\\
&+&\hbar(\Theta(t)-\Theta(t-\tau))(\hat{a}+\hat{a}^\dagger)(\hat{\sigma}_z\gamma+K),\nonumber
\end{eqnarray} 
where $\hat{H}_S$ is the Hamiltonian for the qubit-SQUID oscillator system, $w$ is the qubit energy and $\delta$ the tunneling matrix element, $\hat{H}_B$ is the Hamiltonian for the dissipative environment of the measurement circuitry, $\hat{H}_I$ describes the interaction between the SQUID oscillator and the environment, and  $\Theta$ is the Heaviside step function. We note that for a continuous shape of the current pulse similar results are expected, as long as the switching is not adiabatic. 

Here the SQUID is described, in the lowest-order approximation, by a harmonic oscillator with frequency $\Omega$, i.e.~the plasma frequency of the bias current degree of freedom. This frequency also depends on the applied bias current, as shown in Appendix \ref{parameters}. This dependence leads to a  enhanced ring-down frequency after the pulse is switched off. This change in the SQUID plasma frequency does not, in the first approximation, depend on the qubit state, therefore it will not qualitatively affect this method of discrimination. For the following derivation we assume the SQUID plasma frequency constant (the value during the bias pulse), noting that the ring-down oscillations  occurring in the post-interaction phase have in practice a somewhat higher frequency, but otherwise unchanged behavior.

The dispersive, next-to-leading order component of the qubit-oscillator coupling \cite{measurement07} becomes significant in the absence of a linear component for very weak bias pulse, which is not the limit we investigate here. In the following, the effects of the linear component are investigated. We focus on the regime where the qubit-SQUID interaction displaces the state by more than its zero-point fluctuation but does not yet explore the classical nonlinearity. The first consequence of the nonlinear component may be to add more phase shift to the ringdown oscillations. In the measurement protocol proposed here we assume a symmetric SQUID.

The qubit-oscillator coupling strength is tuned by the bias current $I_B$ \cite{Bertet05d}. When $I_B=0$, the qubit and the SQUID are decoupled. By using a fast current pulse, the qubit-oscillator interaction of arbitrary strength $\gamma$ is turned on only for the short time $\tau$ allowing information about the qubit to be imprinted onto the oscillator.  During this time, the SQUID oscillator is displaced according to the qubit state. After the coupling is switched off, the SQUID oscillator phase particle returns to the original position after  undergoing ring-down oscillations that decay with a damping determined by the SQUID measurement circuitry. The parameter $K$ describes the strength of the bias current kick induced in the oscillator, caused by the abrupt shift in the minimum of the SQUID potential energy from the bias current pulse, in the absence of a qubit.  For the expressions of the parameters $\gamma$ and $K$ and their explicit dependence on $I_B$ see Appendix \ref{parameters}.

During the entire measurement process the oscillator is coupled via a linear Hamiltonian $\hat{H}_I$
\begin{eqnarray}
\hat{H}_I&=&\sum_i\frac{\hbar\lambda_i(\hat{a}\hat{b}^{\dagger}_i+\hat{a}^{\dagger}\hat{b}_i)}{\sqrt{2m\Omega}},
\end{eqnarray}
to a dissipative environment described by a bath of harmonic oscillators
\begin{eqnarray}
\hat{H}_B&=&\sum_i\hbar\omega_i\left(\hat{b}^{\dagger}_i\hat{b}_i+\frac{1}{2}\right),
\end{eqnarray}
with Ohmic spectral density  $J(\omega)=\sum_i\lambda_i^2\hbar\delta(\omega-\omega_i)=m\hbar\kappa\omega\Theta(\omega-\omega_c)/\pi$ \cite{Ingold98}. Here $[\kappa]=s^{-1}$ is the photon loss rate. The cut-off frequency $\omega_c$ is physically motivated by the high-frequency filter introduced by the capacitors. This environment represents the dissipative element contained in any measuring device.

We now describe the dynamics of the qubit and SQUID oscillator during the various phases of our measurement scheme.

\subsection{The interaction phase}

At $t=0$, before the bias current is rapidly pulsed on and the qubit and SQUID interact strongly, we assume the factorized initial state $\hat\rho(0)=\hat\rho_S(0)\otimes\hat\rho_B(0)$. The oscillator interaction with the bath is supposed to be weak, and assuming a Markovian environment, we obtain the standard master equation for the qubit-oscillator reduced density matrix $\hat\rho_S(t)={\rm Tr}_B\left\{\hat\rho(t)\right\}$ in the Born approximation
\begin{eqnarray} 
\dot{\hat{\rho}}_S(t)&=&\frac{1}{\mathbbm{i}
\hbar}\left[\hat{H}_S,\hat{\rho}_S(t)\right] \label{me_bm1}\\
&-&\frac{1}{\hbar^2}\int_0^t
dt'{\rm Tr}_B
\!\left[\hat{H}_{I},\left[\hat{H}_{I}(t,t'),\hat{\rho}_S(t)\otimes\hat{\rho}_B(0)\right]\right],\nonumber 
\end{eqnarray} 
where  
\begin{equation}
\hat{H}_{I}(t,t')= \hat{U}_{t'}^{t}\hat{H}_{I}\hat{U}_{t}^{t'},\:\:\hat{U}_t^{t'}=\mathcal{T}\exp\left(\int_t^{t'}\!\!\!\!d\tau\frac{\hat{H}_S+\hat{H}_B}{\mathbbm{i}
\hbar}\right), 
\end{equation} 
and $\mathcal{T}$ is the time-ordering operator.

This approach is valid at finite temperatures $k_BT \gg  \hbar \kappa$ and times $t\gg1/\omega_c$ \cite{Alicki06,Nato06II}, which is the limit we will discuss henceforth.   

In the qubit $\hat{\sigma}_z$ eigen-basis the density matrix and the qubit-oscillator Hamiltonian read
\begin{eqnarray} 
\hat{\rho}_S & = & \left(\begin{matrix}
\hat{\rho}_{\uparrow\uparrow}&\hat{\rho}_{\uparrow\downarrow} \\
\hat{\rho}_{\downarrow\uparrow}&\hat{\rho}_{\downarrow\downarrow}\end{matrix}\right),\label{matrix}\\
\hat{H}_{S\downarrow\uparrow}&=&\hat{H}_{S\uparrow\downarrow}=\hbar\delta,\:\:\:\:r_{\sigma}=\langle\sigma|\hat{\sigma}_z|\sigma\rangle,\:\:\sigma\in\{\uparrow,\downarrow\},\\
\hat{H}_{S\sigma\sigma}&=&\hbar (r_{\sigma}  w+\Omega(\hat{a}^\dagger
\hat{a}+1/2)\\
&+&(r_{\sigma}\gamma+K)(\hat{a}+\hat{a}^\dagger)).\nonumber
\end{eqnarray} 
In the following, we assume that the environment acts on each matrix element of (\ref{matrix}) in
the same way. This is a valid assumption in the case of very weak damping and $\delta/ w\ll1$ for an Ohmic bath. Within this assumption we obtain 
\begin{eqnarray}
\dot{\hat{\rho}}_{\sigma\sigma}&=&\frac{1}{\mathbbm{i}
\hbar}[\hat{H}_{\sigma\sigma},\hat{\rho}_{\sigma\sigma }]-\mathbbm{i}\delta r_\sigma
(\hat{\rho}_{\downarrow\uparrow}-\hat{\rho}_{\uparrow\downarrow})+\hat{\mathcal{L}}\hat{\rho}_{\sigma\sigma},\\
\dot{\hat{\rho}}_{\uparrow\downarrow}&=&\frac{1}{\mathbbm{i}
\hbar}(\hat{H}_{\uparrow\uparrow}\hat{\rho}_{\uparrow\downarrow}-\hat{\rho}_{\uparrow\downarrow}\hat{H}_{\downarrow\downarrow})+
\mathbbm{i}
\delta(\hat{\rho}_{\uparrow\uparrow}-\hat{\rho}_{\downarrow\downarrow})+\hat{\mathcal{L}}\hat{\rho}_{\uparrow\uparrow}\nonumber\\
\dot{\hat{\rho}}_{\downarrow\uparrow}&=&\frac{1}{\mathbbm{i}
\hbar}(\hat{H}_{\downarrow\downarrow}\hat{\rho}_{\downarrow\uparrow}-\hat{\rho}_{\downarrow\uparrow}\hat{H}_{\uparrow\uparrow})-
\mathbbm{i}
\delta(\hat{\rho}_{\uparrow\uparrow}-\hat{\rho}_{\downarrow\downarrow})+\hat{\mathcal{L}}\hat{\rho}_{\downarrow\uparrow}\nonumber,
\end{eqnarray} 
where 
\begin{eqnarray} 
\hat{\mathcal{L}}\hat{\rho}_{\sigma\sigma'}&=&-\kappa(\hat{a}^\dagger
\hat{a}\hat{\rho}_{\sigma\sigma'}+\hat{\rho}_{\sigma\sigma'} \hat{a}^\dagger
\hat{a}-2\hat{a}\hat{\rho}_{\sigma\sigma'} \hat{a}^\dagger)\\ &-&2\kappa
n(\hat{a}^\dagger \hat{a}\hat{\rho}_{\sigma\sigma'}+\hat{\rho}_{\sigma\sigma'}
\hat{a}\hat{a}^\dagger-\hat{a}\hat{\rho}_{\sigma\sigma'}
\hat{a}^\dagger-\hat{a}^\dagger\hat{\rho}_{\sigma\sigma'} \hat{a}).\nonumber
\end{eqnarray}

At $t=0$ we assume a factorized initial state for the qubit-oscillator reduced density matrix
\begin{equation} 
\hat{\rho}_S(0)=\hat{\rho}_{\rm q}(0)\otimes\hat{\rho}_{\rm HO}(0),
\end{equation} 
and use the Wigner representation of the oscillator density matrix in phase-space \cite{Cahill69}
\begin{eqnarray}
\hat{\rho}_{\rm HO}(0)&=&\frac{1}{\pi}\int\!d^2\!\alpha\:\chi_0(\alpha)\hat{D}(-\alpha),\\
\hat{D}(-\alpha)&=&\exp\left(-\alpha\hat{a}^{\dagger}+\alpha^*\hat{a}\right),
\end{eqnarray} 
where $\chi_0$ is the Fourier transform of the Wigner function. We assume the oscillator to be initially in a thermal state 
\begin{eqnarray} 
\chi_0(\alpha)&=&\frac{1}{4\pi}\exp\left(-\frac{\eta}{2}|\alpha|^2\right),\:\:\eta=1+2n(\Omega),
\end{eqnarray} 
where $n(\Omega)$ is the Bose function at bath temperature $T$. The qubit is assumed to be initially in the pure state $|\Psi\rangle=q_{\uparrow}|\uparrow\rangle+q_{\downarrow}\mathbbm{e}^{\mathbbm
i\phi}|\downarrow\rangle$ such that 
\begin{eqnarray} 
\hat{\rho}_{\rm q}(0)&=&\left(\begin{matrix}q_{\uparrow}^2&q_{\uparrow}q_{\downarrow}\mathbbm{e}^{-\mathbbm{i}\phi}\\ 
q_{\uparrow}q_{\downarrow}\mathbbm{e}^{\mathbbm{i}\phi}& q_{\downarrow}^2\end{matrix}\right).
\end{eqnarray}

For the corresponding Wigner characteristic functions we obtain the following coupled partial differential
equations:
\begin{eqnarray} 
\dot{\chi}_{\sigma\sigma}&=&(\mathbbm{i}(r_{\sigma}\gamma+K)(\alpha+\alpha^*)+\mathbbm{i}
\Omega(\alpha\partial_{\alpha}-\alpha^*\partial_{\alpha^*})\nonumber\\
&+&\mathcal{D})\chi_{\sigma\sigma}-r_{\sigma}\mathbbm{i}
{\delta}(\chi_{\downarrow\uparrow}-\chi_{\uparrow\downarrow})),\label{ugly_eqs}\\
\dot{\chi}_{\uparrow\downarrow}&=&(2\mathbbm{i}
\gamma(\partial_{\alpha^*}-\partial_{\alpha})+\mathbbm{i}
\Omega(\alpha\partial_{\alpha}-\alpha^*\partial_{\alpha^*})-2\mathbbm{i}
 w\nonumber\\
&+&\mathbbm{i}K(\alpha+\alpha^*)+\mathcal{D})\chi_{\uparrow\downarrow}-\mathbbm{i}
{\delta}(\chi_{\downarrow\downarrow}-\chi_{\uparrow\uparrow})),\nonumber\\
\dot{\chi}_{\downarrow\uparrow}&=&(-2\mathbbm{i}
\gamma(\partial_{\alpha^*}-\partial_{\alpha})+\mathbbm{i}
\Omega(\alpha\partial_{\alpha}-\alpha^*\partial_{\alpha^*})+2\mathbbm{i}
 w \nonumber\\ 
&+&\mathbbm{i}K(\alpha+\alpha^*)+\mathcal{D})\chi_{\downarrow\uparrow}+\mathbbm{i}
{\delta}(\chi_{\downarrow\downarrow}-\chi_{\uparrow\uparrow})),\nonumber
\end{eqnarray} 
where the differential operator $\mathcal{D}$ is given by
\begin{eqnarray}
\mathcal{D}&=&-\kappa(\alpha\partial_{\alpha}+\alpha^*\partial_{\alpha^*})-\eta\kappa
|\alpha |^2.
\end{eqnarray}
To solve these equations, we approximate the inhomogeneous parts, in the limit  of short time $\tau$ and weak tunneling $\delta$, by
\begin{eqnarray} 
\chi_{\sigma\sigma'}(t)\simeq\chi_{\sigma\sigma'}(0)+t\dot{\chi}_{\sigma\sigma'}(0),\:\:\:\sigma,\sigma'\in\{\uparrow,\downarrow\} .\label{important}
\end{eqnarray}
For details on the solution see Appendix \ref{solution_chi}.
\subsection{The post-interaction phase}
The state prepared by the interaction with the qubit at $t=\tau$, as the bias current pulse ends, is described by
\begin{eqnarray}
\hat\rho(\tau)&=&\sum_{\sigma,\sigma'\in\{\uparrow,\downarrow\}}|\sigma\rangle\langle\sigma'|\hat\rho_{\sigma\sigma'}(\tau)\otimes\hat\rho_B(0).
\end{eqnarray}

Since the system Hamiltonian no longer contains any qubit-oscillator interaction, we can write the time evolution of this density matrix as follows 
\begin{eqnarray}
\hat\rho(t)&=&\!\!\!\!\sum_{\sigma,\sigma'\in\{\uparrow,\downarrow\}}\!\!\!\!
\hat U_{\rm q}(t)|\sigma\rangle\langle\sigma'|\hat U^{\dagger}_{\rm q}(t)\nonumber\\
&\cdot&\hat U_{{\rm HO}-B}(t)\hat\rho_{\sigma\sigma'}(\tau)\otimes\hat\rho_B(0)\hat U^{\dagger}_{{\rm HO}-B}(t),\label{ref6a}
\end{eqnarray}
where the evolution operators are given by
\begin{eqnarray}
\hat U_{\rm q}(t)&=&\exp(-\mathbbm{i}(t-\tau)(\delta\hat\sigma_x+ w \hat\sigma_z)),\\
\hat U_{{\rm HO}-B}&=&\mathcal{T}\exp\left(\int_\tau^t\! dt'\:\frac{\hat H_B+\hat H_I+\hbar\Omega\hat a^\dagger\hat a)}{\mathbbm{i}\hbar}\right).
\end{eqnarray}
In the reduced density matrix
\begin{eqnarray}
\hat \rho_S(t)&=&{\rm Tr}_B\hat\rho(t)=\!\!\!\!\sum_{\sigma,\sigma'\in\{\uparrow,\downarrow\}}\!\!\!\!
\hat U_{\rm q}(t)|\sigma\rangle\langle\sigma'|\hat U^{\dagger}_{\rm q}(t)\\
&\cdot&{\rm Tr}_B\left\{\hat U_{{\rm HO}-B}(t)\hat\rho_{\sigma\sigma'}(\tau)\otimes\hat\rho_B(0)\hat U^{\dagger}_{{\rm HO}-B}(t)\right\},\nonumber
\end{eqnarray}
we can treat the time evolution of the oscillator components by means of a master equation in the Born-Markov approximation and,
in a similar manner to Eq.~(\ref{me_bm1}), we obtain
\begin{eqnarray} 
\dot{\hat{\rho}}_{\sigma\sigma'}(t)&=&-\mathbbm{i}\Omega[\hat a^\dagger \hat a,\hat{\rho}_{\sigma\sigma'}(t)]\label{ref6b}\\
&-&\frac{1}{\hbar^2}\int_0^\infty
\!\!\!\!dt'{\rm Tr}_B
\!\left[\hat{H}_{I},[\hat{H}_{I}(t,t'),\hat{\rho}_{\sigma\sigma'}(t)\otimes\hat{\rho}_B(0)]\right].\nonumber 
\end{eqnarray} 
Using the Wigner representation
\begin{eqnarray} 
\hat{\rho}_{\sigma\sigma'}(t)&=&\frac{1}{\pi}\int\!d^2\alpha\:\tilde\chi_{\sigma\sigma'}(\alpha,t)\hat{D}(-\alpha),
\end{eqnarray} 
we obtain the differential equation 
\begin{eqnarray} 
\dot{\tilde{\chi}}_{\sigma\sigma'}(\alpha, t)&=&(\mathbbm{i}
\Omega(\alpha\partial_{\alpha}-\alpha^*\partial_{\alpha^*})+\mathcal{D})\tilde{\chi}_{\sigma\sigma'}(\alpha,t),
\end{eqnarray} 
with the initial condition prepared at the end of the interaction phase
\begin{eqnarray} 
\tilde\chi_{\sigma\sigma'}(\alpha,\tau)=\chi_{\sigma\sigma'}(\alpha,\tau),
\end{eqnarray} 
and the analytic solution
\begin{eqnarray} 
\tilde{\chi}_{\sigma\sigma'}(\alpha,t)&=&\tilde{\chi}_{\sigma\sigma'}(\alpha
\mathbbm{e}^{-(t-\tau)(\kappa-\mathbbm{i}
\Omega)},\tau)\nonumber\\
&\times&\exp\left(\frac{\eta}{2}|\alpha|^2(\mathbbm{e}^{-2(t-\tau)\kappa}-1)\right).
\end{eqnarray} 
The reduced density matrix in the post-interaction phase is given by
\begin{eqnarray}
\hat\rho_S(t)&=&\!\!\!\!\!\sum_{s,s'\in\{\uparrow,\downarrow\}}\!\!\!\!
|s\rangle\langle s'|\frac{1}{\pi}\int
d^2\alpha \chi_{ss'}(\alpha,t)\hat{D}(-\alpha),
\end{eqnarray}
where
\begin{eqnarray}
\chi_{ss'}(\alpha,t)\!\!&=&\!\!\!\!\!\!\!\!\!\!\sum_{\sigma,\sigma'\in\{\uparrow,\downarrow\}}\!\!\!\!\!\!\langle s|\hat U_{\rm q}(t)|\sigma\rangle\langle \sigma'|\hat U_{\rm q}^\dagger(t)|s'\rangle\tilde\chi_{\sigma\sigma'}(\alpha,t)\label{rel1}.
\end{eqnarray}
In the the post-interaction phase, the qubit and the oscillator are decoupled. The trace of the oscillator-bath part in Eq.~(\ref{ref6a}) is time independent, as one can see after a circular permutation of the involved operators. One finds that the qubit time evolution is given only by the unitary $\hat U_q$, and thus is independent of the oscillator. Physically, this means that in the post-interaction phase no further information about the qubit can be transferred to the oscillator-bath system, and thus the qubit suffers no further decoherence. 

\section{Results}\label{results}

In this section we analyze the qubit decoherence and the evolution of its detector, the dissipative oscillator, during the entire measurement process.

\subsection{Qubit decoherence} 
During the interaction phase, $t\in(0,\tau)$, the qubit is in contact with an environment represented by the dissipative oscillator, and  thus subject to decoherence. 

The qubit can be prepared in a well defined state by thermal relaxation or (if the temperature is too high) by measurement post-selection and conditional rotation by microwave pulses. 

\begin{figure}[!h]
\includegraphics[width=\columnwidth]{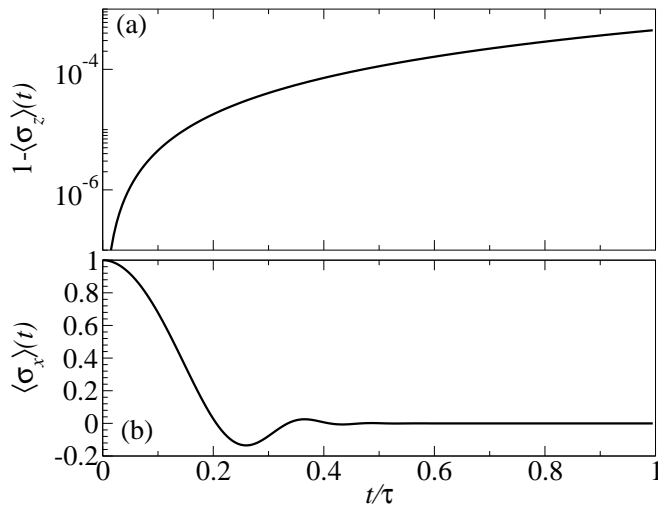} 
 \caption{
(a) Evolution of $\langle\hat \sigma_z\rangle$ with qubit initially in $|\uparrow\rangle$ state. 
(b) Dephasing from the $1/\sqrt{2}(|\uparrow\rangle+|\downarrow\rangle)$ state
for the time $\tau$ that the qubit is in contact with the oscillator.
For both plots, the following parameters were used: 
$\Omega/(2\pi)=0.97$ GHz, $\kappa/\Omega=10^{-2}$, $ w =\Omega$, $\Omega\tau=1.83$, $\delta\tau=0.015$, $\gamma\tau=3$, $T=30$ mK. The assumed values of the circuit parameters are given in  Appendix \ref{parameters}.}\label{decoherence}
\end{figure}

We analyze the qubit relaxation described by  
\begin{equation}
\langle\hat{\sigma}_z\rangle(t)=4\pi(\chi_{\uparrow\uparrow}(0,t)-\chi_{\downarrow\downarrow}(0,t))
\end{equation}
and from Eq.~(\ref{chi_d}) we obtain the analytic result 
\begin{eqnarray}
\langle\hat{\sigma}_z\rangle(t)&=&(q_{\uparrow}^2-q_{\downarrow}^2)(1- 2t^2\delta^2)\\
&+&4q_{\uparrow}q_{\downarrow}t\delta(t w \cos(\phi)+\sin(\phi))\nonumber. 
\end{eqnarray}
We observe that the above expression is identical with the expansion up to the second order in time of $\langle\hat{\sigma}_z\rangle(t)$ when the qubit evolves under the free Hamiltonian $\hat H_{\rm q}$ only. Thus, the evolution of $\langle\hat{\sigma}_z\rangle(t)$ in this short time expansion is indistinguishable from the free evolution of the unperturbed qubit. This can be understood as follows: the observable $\hat \sigma_z$ commutes with the environment coupling, but is not an integral of the free motion, as required for a QND measurement \cite{Braginsky95}. Thus, the perturbation of the measured observable comes only from the free evolution of the system. One can restrict this perturbation by reducing the time $\tau$ when it takes place.  Fig.~\ref{decoherence} (a) shows the evolution of $\langle\hat{\sigma}_z\rangle(t)$ for a set of parameters closely related to a feasible experiment, see also Appendix \ref{parameters}. The initial state chosen for panel (a) was $|\uparrow\rangle$.

Furthermore, we analyze the qubit coherence $\langle\hat{\sigma}_x\rangle$ which is given by
\begin{eqnarray}
\langle\hat{\sigma}_x\rangle(t)&=&8\pi{\rm Re}\chi_{\uparrow\downarrow}(0,t),
\end{eqnarray} 
and can be evaluated from Eqs.~(\ref{chi_off1},\ref{chi_off2}), where $\chi_{\uparrow\downarrow}^{\rm inh}(0,t)$ can be integrated numerically. 

We observe that, if the interaction time $\tau$ is long enough to allow the oscillator a full period evolution, one finds a revival in the qubit coherence at the end of this period. As the oscillator returns to (almost) its initial state, the information about the qubit is ``erased" from the oscillator, as the oscillator states corresponding to $|\uparrow\rangle$ and $|\downarrow\rangle$ are no longer discernible. The height of the coherence revival peaks at $\Omega t=2\pi n$ decays in time as the information about the coupled qubit-oscillator system flows (irreversibly in this case) into the environment. 

The qubit dephasing for the same parameters of Appendix \ref{parameters} is shown in Fig.~\ref{decoherence} (b). The appropriate initial state  for this study is the equal superposition $(1/\sqrt{2})(|\uparrow\rangle+|\downarrow\rangle)$. We observe that the qubit appears completely dephased after the strong interaction with the
damped oscillator, such that only a classical probability is imprinted
onto the latter.  

In Fig.~\ref{decoherence} (a) we observe that the relaxation from the excited qubit state is very weak during the interaction time, as $\langle\hat\sigma_z\rangle$ differs at most by $10^{-3}$  from the initial value of 1. This combination of low coherence (b), indicating the fact that the information about the qubit has been imprinted onto the oscillator, and very low relaxation (a) demonstrates that the first step of the measurement protocol produces a good starting point for the second one, the oscillator readout. The negligible relaxation brings the qubit close to QND dynamics.

We observe that the qubit coherence time is essentially dominated by the coupling between the qubit and its complex environment $\gamma^{-1}$ such that it is desirable to achieve $\gamma\tau\gg 1$.  The relaxation of the qubit has been described in the first order in time, and essential to the almost-QND result is that $\tau\delta\ll1$. We note that the implied condition $\gamma\ll\delta$ contradicts none of our approximations, and can also be realized in experiment.
\subsection{Detector dynamics}
In this section we study the evolution of the damped oscillator, which represents the detector. To achieve the strong qubit-oscillator coupling during the short interaction phase required to imprint the qubit state onto the oscillator, one needs a  bias current pulse that approaches the critical current for the SQUID. Nonetheless, it is important that the SQUID does not switch out to the running state during the bias current pulse. For the parameters given in Appendix \ref{parameters}, we can evaluate the SQUID escape rate \cite{Martinis87} from the zero-voltage state during the bias current pulse in the regime of quantum assisted thermal activation $(k_BT\apprge\hbar\Omega)$ 
\begin{eqnarray}
\Gamma_{\rm sw}&=&\frac{\sinh\left(\frac{\hbar\Omega}{2k_BT}\right)}{\sin\left(\frac{\hbar\Omega}{2k_BT}\right)}\frac{\Omega}{2\pi}\exp\left(\frac{-\Delta U}{k_BT}\right),
\end{eqnarray} 
where $\Delta U$ is the potential barrier. We obtain, for the worst case, $\Gamma_{\rm sw}\approx 3.6\cdot 10^{7}s^{-1}$  such that the escape time is much larger than the duration of the bias current  pulse.

The output of the detector is the time dependent voltage across the SQUID, which is proportional 
to the momentum of the oscillator. The probability distribution of momentum is given by
\begin{eqnarray}
P(p,\tau,t)&=&\mu\langle\delta(\hat{p}-p)\rangle\label{prob_dist}\\
&=& 2 \int d
\alpha_x\sum_{\sigma\in\{\uparrow,\downarrow\}}\chi_{\sigma\sigma}(\alpha_x,t)\exp\left(\frac{\mathbbm{i} p \alpha_x}{\mu}\right),\nonumber\\
\mu&=&\sqrt{\frac{m\Omega\hbar}{2}},\:\:\alpha=\alpha_x+\mathbbm{i}\alpha_y,
 \end{eqnarray}
where, in the post-interaction phase $(t>\tau)$, $\chi_{\sigma\sigma}(\alpha_x,t)$ also depends on $\tau$ via its initial condition. The expectation values for the ${\rm n}^{th}$ moment of the oscillator momentum and position are then 
\begin{eqnarray} 
\langle\hat{p}^n\rangle(t)&=&\frac{4\pi\mu^n}{\mathbbm{i}^n}(-1)^n(\partial_{\alpha_x})^n\!\!\!\!\sum_{\sigma\in\{\uparrow,\downarrow\}}\!\!\!\chi_{\sigma\sigma}(\alpha_x,t)|_{\alpha_x=0}\label{momentum},\\
\langle\hat{x}^n\rangle(t)&=&\!\!\left(\!\!\sqrt{\frac{\hbar}{2m\Omega}}\!\right)^n\!\!\frac{4\pi}{\mathbbm{i}^n}(\partial_{\alpha_y})^n\!\!\!\!\sum_{\sigma\in\{\uparrow,\downarrow\}}\!\!\!\chi_{\sigma\sigma}(\mathbbm{i}\alpha_y,t)|_{\alpha_y=0}.\nonumber
\end{eqnarray} 
Furthermore, in the post-interaction phase we have, from Eq.~(\ref{rel1}),
\begin{eqnarray} 
\!\!\!\sum_{\sigma\in\{\uparrow,\downarrow\}}\!\!\!\!\chi_{\sigma\sigma}&=&\!\!\!\!
\!\!\!\!\sum_{s,s',\sigma\in\{\uparrow,\downarrow\}}\langle\sigma|\hat U_{\rm q}(t)|s\rangle\langle s'|\hat U^\dagger_{\rm q}(t)|\sigma\rangle\tilde\chi_{ss'}\nonumber\\
&=&\sum_{s,s'}\langle s'|\hat U^\dagger_{\rm q}(t)\hat U_{\rm q}(t)|s\rangle\tilde\chi_{ss'}=\sum_{s}\tilde\chi_{ss},
\end{eqnarray} 
which shows, as expected, that no measurement of the oscillator can provide information about the post-interaction evolution of the qubit, provided this evolution is unitary (i.e.~the qubit is not being measured by something else).

For the evaluation of both Eqs.~(\ref{prob_dist}, \ref{momentum}) the $s$-integration in $\chi_{\sigma\sigma}^{\rm inh}$, Eq.~(\ref{chi_diag}) should be evaluated last. Thus, one obtains an analytic (but rather long) expression for the expectation value of momentum, while for the
probability density a numerical $s$-integration is required.  Nevertheless, the components originating in $\chi_{\sigma\sigma}^{\rm hom}$ turn out to be dominant, and we give their
analytic expressions in the following:
 \begin{eqnarray} 
&&\langle\hat p\rangle(\tau,t)=\langle\hat p\rangle_{\rm hom}(\tau,t)
 						+\langle\hat p\rangle_{\rm inh}(\tau,t),\label{mom_expect}\\ 
&&\langle\hat p\rangle_{\rm hom}=\left(K+\gamma q_{\uparrow}^2-\gamma q_{\downarrow}^2\right) \mu\mathbbm{e}^{-(t-\tau)\kappa}  
\nonumber\\
&\cdot&\left(\mathbbm{e}^{-(t-\tau)\mathbbm{i} \Omega}\frac{1-\mathbbm{e}^{-\tau (\kappa +\mathbbm{i} \Omega)}}{-\kappa -\mathbbm{i} \Omega }
+\mathbbm{e}^{(t-\tau)\mathbbm{i} \Omega}\frac{1-\mathbbm{e}^{-\tau(\kappa-\mathbbm{i} \Omega)}}{-\kappa +\mathbbm{i}\Omega }\right)\nonumber,
\end{eqnarray} 
The explicit form of the probability distribution of momentum, Eq.(\ref{prob_dist}), is given by 
\begin{eqnarray} 
P(p,\tau,t)&=&P_{\rm hom}(p,\tau,t)+P_{\rm inh}(p,\tau,t),\label{prob_dist_hom}
\end{eqnarray} where
\begin{eqnarray*} 
P_{\rm hom}(p,\tau,t)=
\sum_\sigma\frac{|\langle\sigma|\Psi\rangle|^2}{\sqrt{2\pi\eta}}
\exp\left(\!\!\frac{\mathbbm{i}p}{\sqrt{2\eta}\mu}-\mathbbm{i}\frac{K+r_\sigma \gamma}{\sqrt{2\eta}}
\mathbbm{e}^{(\tau-t)\kappa} \right. 
&&\nonumber\\
\cdot\!\!\left.\left(\mathbbm{e}^{-(t-\tau)\mathbbm{i} \Omega}\frac{1-\mathbbm{e}^{-\tau (\kappa +\mathbbm{i} \Omega)}}{-\kappa -\mathbbm{i} \Omega }
+\mathbbm{e}^{(t-\tau)\mathbbm{i} \Omega}\frac{1-\mathbbm{e}^{-\tau(\kappa-\mathbbm{i} \Omega)}}{-\kappa +\mathbbm{i}\Omega }\right)
\!\!\right)^2.&&
\end{eqnarray*}
The results above refer to the post-interaction phase $t>\tau$.  For the interaction phase, $t\in(0,\tau)$, the probability distribution of momentum is given by $P(p,t,t)$ in Eq.~(\ref{prob_dist_hom}) and the expectation value of momentum by $\langle \hat p\rangle(t,t)$ in Eq.~(\ref{mom_expect}), i.e.~by replacing $\tau$ by $t$.

The expectation value of momentum $\langle \hat p\rangle(\tau,t)$ 
in the post-interaction phase contains information about the qubit initial state.  We observe that the momentum oscillations corresponding to the two different initial qubit states $|\uparrow\rangle$ and $|\downarrow\rangle$ for $t>\tau$ are in phase. Disregarding the inhomogeneous contributions, which are relatively small in the limit of small $\tau\delta$, the envelope of the homogeneous part is given by
 \begin{eqnarray} 
 \mathcal{A}(q_{\uparrow},q_{\downarrow})&=&\frac{2\mathbbm{e}^{-(t-\tau) \kappa }(K+\gamma q_{\uparrow}^2-\gamma q_{\downarrow}^2)  \mu }{\sqrt{\kappa ^2+\Omega ^2}}\nonumber\\
&\cdot&\sqrt{-2
\mathbbm{e}^{-\kappa  \tau } \cos (\tau  \Omega )+\mathbbm{e}^{-2 \kappa  \tau
}+1}.\label{amplitude}
 \end{eqnarray}  

\begin{figure}[!h]
    \includegraphics[width=0.48\textwidth]{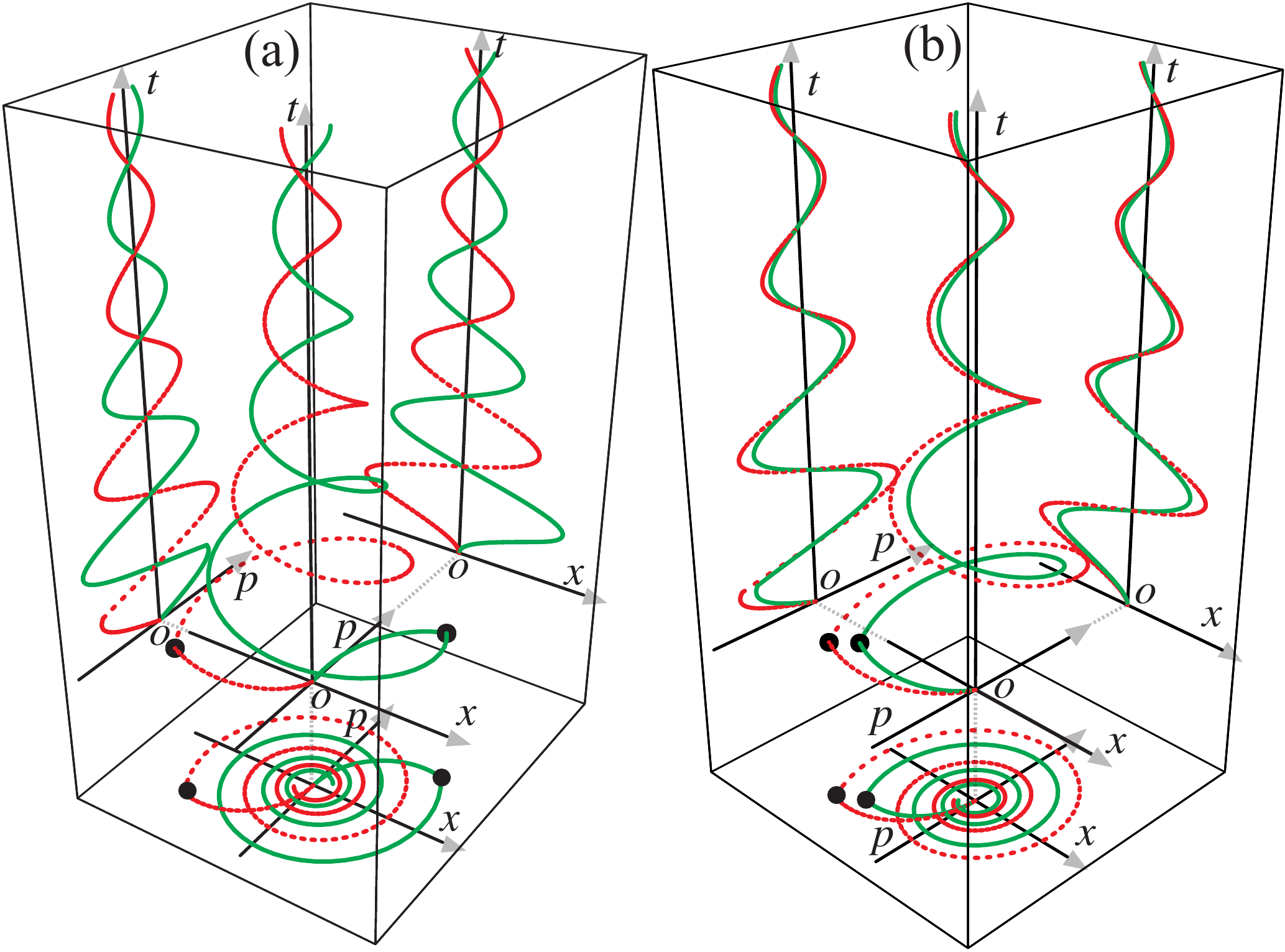}  
    \caption{(Color online). Phase space representation of the oscillator trajectories $(\langle \hat x\rangle(t),\langle \hat p\rangle(t),t)$ corresponding to the two qubit states $|\downarrow\rangle$ (dashed, red) and $|\uparrow\rangle$
(continuous, green) for the parameters given in Appendix \ref{parameters}, an oscillator quality factor of $10$, with $K=0$ (a) and $K\not =0$ (b). Projections on the $(x,p)$, $(x,t)$ and $(p,t)$ planes are included. Both trajectories start at the origin and move away from it under the influence of the interaction with the qubit. At the point marked with $\bullet$ the interaction is switched off, and the system evolves freely spiraling around the origin. The trajectories circle around each other without crossing. }\label{spiral}
 \end{figure}

Fig.~\ref{spiral} illustrates the phase-space trajectories of the oscillator corresponding to the qubit being in either the $|\uparrow\rangle$ or $|\downarrow\rangle$ state. During the interaction phase the system moves away from the origin. After switching off the interaction, the trajectories spiral back towards the origin, without crossing. For $K=0$ the trajectories are symmetric with respect to the origin, while $K\not=0$ introduces an asymmetry. We note that the artificial situation $K=0$ includes only the bare oscillator response for the 
different qubit states. This situation has been introduced in order to more easily illustrate the difference between the two oscillations.

\begin{figure}[!h]
  \includegraphics[width=\columnwidth]{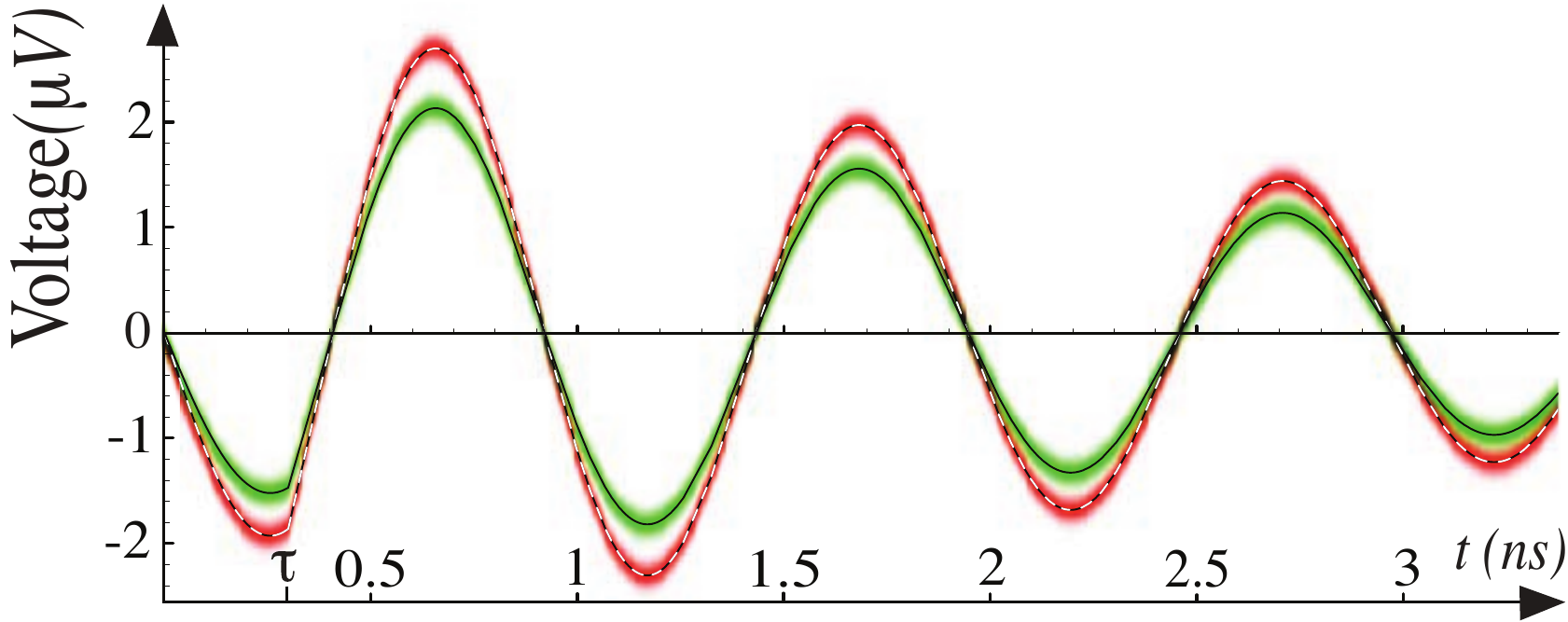}
     \caption{(Color online) Probability distribution of output voltage (density plot,
dark color indicates high and white low density) and expectation value of momentum for the two qubit states $|\downarrow\rangle$ (dashed, red) and $|\uparrow\rangle$
(continuous, green). Here $\Omega/(2\pi)=0.97$ GHz, $\Omega/\kappa=20$, $ w =\Omega$, $\Omega\tau=1.83$, $\delta\tau=0.015$, $\gamma\tau=3$, $T=30$ mK. The assumed values of the circuit parameters are given in  Appendix \ref{parameters}.}\label{detection}
 \end{figure}
Fig.~\ref{detection} shows the output of the detector for the two qubit states $|\downarrow\rangle$ and $|\uparrow\rangle$. 

The standard condition for the possibility of single-shot readout, i.~e.,~the maximal separation of the two peaks corresponding to different qubit states in the probability distribution Eq.~(\ref{prob_dist}) should be larger than the peak width, is given by
\begin{equation} 
 \varepsilon\approx\frac{|\mathcal{A}(1,0)-\mathcal{A}(0,1)|}{3\mu\sqrt{\eta}}>1,\label{single_shot}
 \end{equation}  
where the envelope (\ref{amplitude}) has been evaluated at $t=\tau$. We note that $q_{\uparrow}$ and $q_{\downarrow}$ are continuous variables with values between $0$ and $1$ and the condition presented above takes into account the extremal case of the difference between the states $|\uparrow\rangle$ and $|\downarrow\rangle$. The result is independent of $K$. For the parameters of Fig.~\ref{detection} we have $\varepsilon\approx2.5$.
 
\section{Practical implementation}\label{implementation}
\begin{figure}[!h]
  \includegraphics[width=\columnwidth]{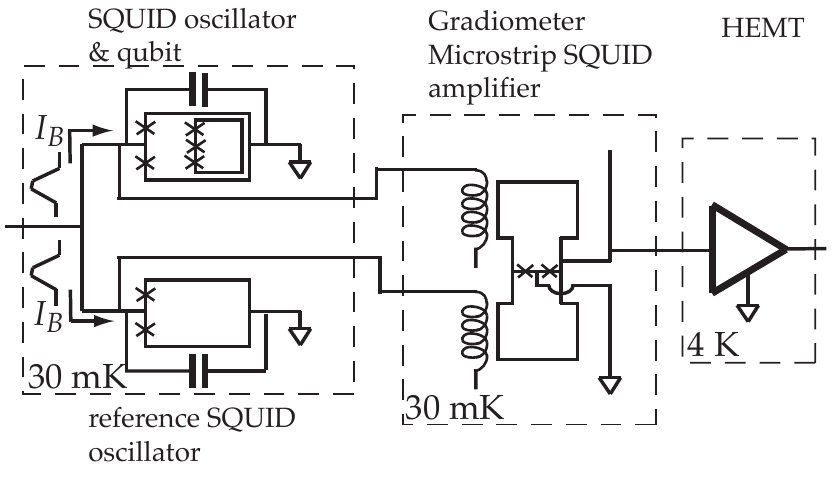}
\caption{Circuit diagram for SQUID oscillator and qubit, along with reference SQUID oscillator, dual-input gradiometer microstrip amplifier and a cryogenic High Electron Mobility Transistor (HEMT). Dashed boxes indicate different chips and/or different temperatures.}\label{implement}
 \end{figure}

A possible measurement protocol involves discriminating the amplitudes of the ringdown oscillations corresponding to different qubit states. As demonstrated by Eq.~(\ref{amplitude}), the amplitude difference is independent of $K$. This discrimination could be performed more accurately with an interferometric technique, where ringdown oscillations from a second, reference SQUID oscillator that is not coupled to the qubit are combined with those from the original SQUID oscillator.The reference SQUID is biased such that it undergoes ringdown oscillations with the same phase and amplitude as those of the measurement SQUID oscillator for one of the two qubit states. In this case, the resultant signal after the subtraction would be exactly zero for perfect cancellation when the qubit state causes the two SQUID oscillators to have identical ringdown signals. A residual ringdown oscillation would be produced for the other qubit state. 
This scheme requires that the two SQUIDs receive an identical kick and begin their ringdown oscillations at the same time. This can be achieved by splitting the bias current pulse signal along two separate lines, one going to each SQUID, as shown in Fig.~\ref{implement}, where the layout is such that the reference SQUID has a vanishing coupling to the qubit. 

\begin{figure}[!h]
  \includegraphics[width=\columnwidth]{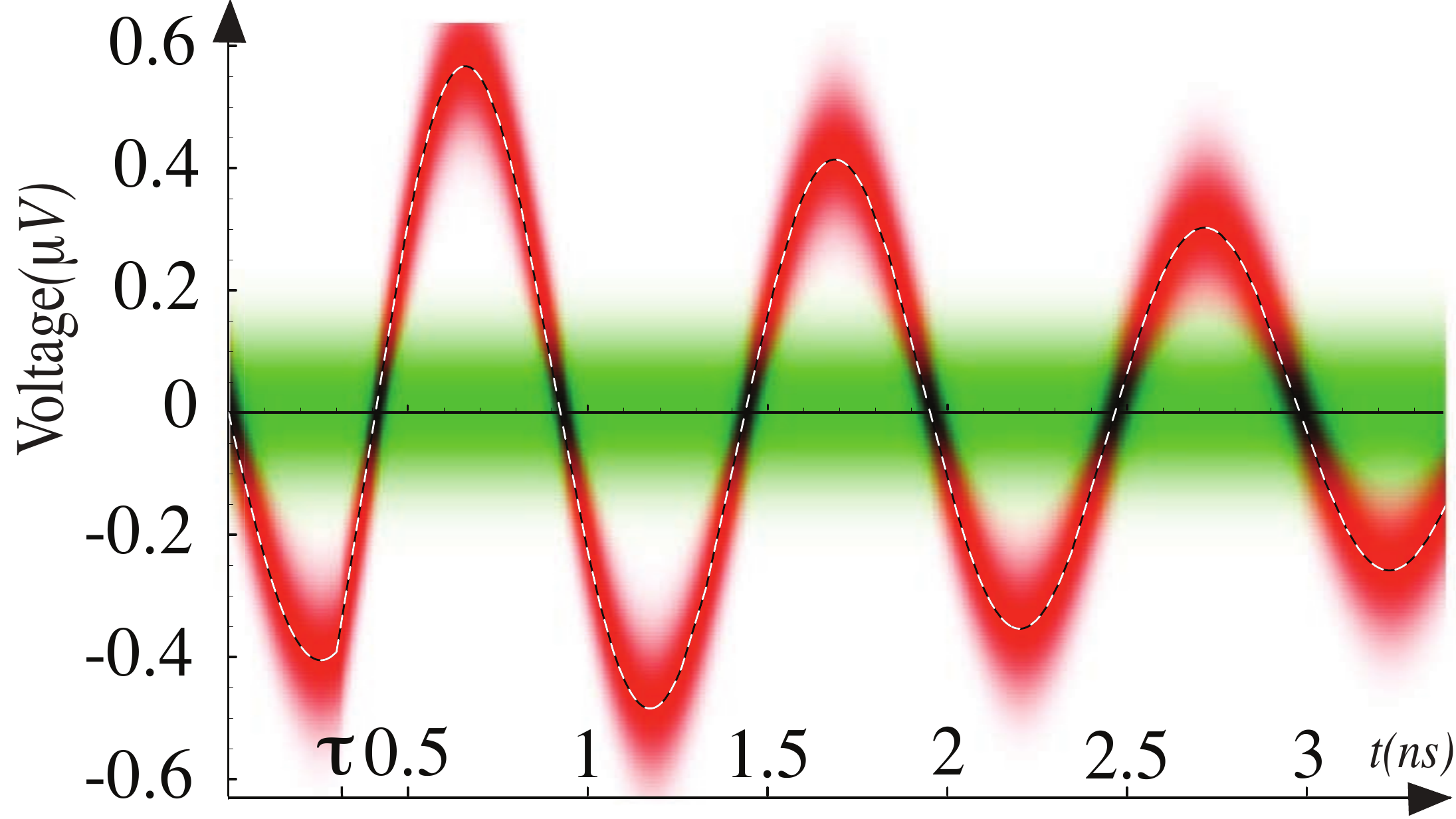}
     \caption{(Color online) Probability distribution of output voltage (density plot,
dark color indicates high and white low density) and expectation value of momentum for the two qubit states $|\downarrow\rangle$ (dashed, red) and $|\uparrow\rangle$
(continuous, green). Here the contribution of the reference SQUID has been introduced. Parameters: $\Omega/(2\pi)=0.97$ GHz, $\Omega/\kappa=20$, $ w =\Omega$, $\Omega\tau=1.83$, $\delta\tau=0.015$, $\gamma\tau=3$, $T=30$ mK. The assumed values of the circuit parameters are given in  Appendix \ref{parameters}.\label{detection2}}
 \end{figure}

Fig.~\ref{detection2} shows the total signal, i.e.~the difference of the ringdown oscillations from the measurement and reference SQUIDs for the two qubit states. We have considered the case where the total flux bias for the reference SQUID is equal to the total flux bias for the measurement SQUID in the case where the qubit state is $|\uparrow\rangle$. In this case the difference signal is smeared around  $0$ for the qubit in state $|\uparrow\rangle$. If the qubit is in the $|\downarrow\rangle$ state, the output signal oscillates with an amplitude is given by the difference between the two ringdown oscillations in Fig.~\ref{detection}.
 
The subtraction of the two ringdown signals can be achieved by using a microstrip SQUID amplifier arranged as a gradiometer with two separate microstrip inputs with their senses indicated in Fig.~\ref{implement} \cite{Mueck08}. 
The microstrip SQUID amplifier consists of a dc SQUID with a multi-turn superconducting input coil above a conventional SQUID washer, where the signal is connected between one side of the input coil and ground and the other end of the input coil is left open. Input signals near the stripline resonance frequency, related to the total length of the input coil, typically of the order of $1$~GHz, couple strongly to the SQUID loop and the SQUID produces an output signal with a gain of $\sim 10-20$~dB \cite{Mueck03}.
A gradiometer microstrip SQUID amplifier for amplifying the difference between two separate signals near the stripline resonance can be produced as a straightforward extension from previous microstrip SQUID layouts by using a SQUID geometry with two loops and a separate stripline coil coupled to each of the loops, with one signal input connected to each stripline \cite{Mueck08}.
With no crosstalk between the two inputs, the circulating currents in the two loops of the SQUID amplifier cancel out when the input signals are identical, resulting in a vanishing output signal. Thus, with the arrangement in Fig.~\ref{implement}, the microstrip SQUID amplifier produces the difference between the two oscillator ringdowns. Of course, in any practical gradiometer, there will be non-zero crosstalk, where a signal at one input induces circulating currents in the other loop of the SQUID amplifier. However, for reasonable layouts of the device, this crosstalk could be kept at the $1$\% level, thus setting a limit on the fidelity of the subtraction \cite{Mueck08}.

Based on the calculated difference signals for the ringdown oscillations in the two qubit states from Fig.~\ref{detection2}, one must be able to discriminate the oscillations for the $|\downarrow\rangle$ qubit state from the non-oscillatory signal for the $|\uparrow\rangle$ state. Thus one needs to resolve a $\sim 1$~GHz signal with an amplitude of $\sim 0.5$~$\mu$V in a $\sim 100$~MHz bandwidth, i.e.~before the ringdown is completed. Microstrip SQUID amplifiers operated at $20$~mK have achieved noise temperatures as low as $\sim 50$~mK \cite{Mueck01}. If we assume a conservative noise temperature estimate of $200$~mK for our gradiometer microstrip SQUID amplifier, this would correspond to a noise of $250$~ nV in the $100$~MHz bandwidth referred back to the SQUID oscillators. Thus, it should be possible to discriminate between the two possible output signals corresponding to the two qubit states in a single shot. 

In the non-ideal case, the noise of the reference SQUID increases the broadening of the curves in Fig.~\ref{detection2} such that the single shot condition (\ref{single_shot}) must accommodate another width $\eta$. Still, at the parameters used in Fig.~\ref{detection2}, this condition will still hold.

\section{Conclusion} 
We have demonstrated that a non-QND
Hamiltonian can induce a close to QND backaction on the qubit, despite
arbitrarily strong interaction with the environment, provided that the
interaction time is very short, i.e.~the measurement is
quasi-instantaneous.  The relaxation of the qubit has been described
in the first order in time and, essential to the almost-QND results
presented above, is that $\tau\delta\ll1$.

We observe that the measurement time, i.e.,~the time needed to reduce
the qubit density matrix to a classical mixture is essentially
dominated by the coupling between the qubit and its complex
environment $\gamma^{-1}$ such that it is desirable to achieve
$\gamma\tau\gg 1$. 

The readout time for the oscillator is restricted only by the
ring-down of the two possible oscillations of momentum,
i.e.~$\kappa^{-1}$. The amplitude of these oscillations is
proportional to $\gamma$, which again stresses the usefulness of a
strong qubit-oscillator coupling.  If the two peaks in $P(p,\tau,t)$ become
separated by significantly more than their widths, single shot
measurement may become possible.

The method presented above has the advantage of a very short interaction between the qubit and its environment, compared to e.g.~the dispersive readout of Ref.~\cite{Lupascu04}, and results in a QND-type of readout, without the requirement of strong, continuous AC driving of e.g.~Ref.~\cite{Lupascu07} which may induce spurious qubit relaxation. 

As a figure of merit we consider the QND fidelity in Ref.~\cite{ralph06}. For the parameters used in Fig.~\ref{decoherence} and an initial qubit state $\Psi=1/\sqrt{2}(|\uparrow\rangle+|\downarrow\rangle)$, our scheme achieves at the end of the post interaction phase a QND fidelity of $99.92\%$.

Furthermore, if the aim is to apply the idea of a short interaction with an intermediate system, dispersive measurement, with all its potential advantages, may be difficult due to the continuous driving which implies continuous interaction between the qubit and its environment.

\section{Acknowledgment}
We acknowledge useful discussions with Michael M\"{u}ck. 
This work was supported by DFG through SFB 631, by 
NSERC discovery grants, by QuanumWorks and by EU through EuroSQIP. 
\appendix
\section{Solution for the Wigner characteristic functions}\label{solution_chi}
In this section we solve Eqs.~(\ref{ugly_eqs}) using the approximation (\ref{important}).
\subsubsection{The diagonal density matrix elements}
We solve the diagonal equations needed for evaluation of expectation values such as $\langle \hat p\rangle(t)$, which characterize the output of the detector:
\begin{eqnarray}
 \dot{\chi}_{\sigma\sigma}&=&(\mathbbm{i}(r_{\sigma}\gamma+K)(\alpha+\alpha^*)
 +\mathbbm{i}\Omega(\alpha\partial_{\alpha}-\alpha^*\partial_{\alpha^*})\nonumber\\
&+&\mathcal{D})\chi_{\sigma\sigma}-r_\sigma\mathbbm{i} {\delta}\chi_0(\alpha)F(\alpha,t),\label{eq:1}
\end{eqnarray} 
where
\begin{eqnarray} 
F(\alpha,t)&=&2q_{\uparrow}q_{\downarrow}\sin(\phi)(\mathbbm{i}-K(\alpha+\alpha^*)t)-2\mathbbm{i}(q_{\uparrow}^2-q_{\downarrow}^2)\delta t\nonumber\\
&-&2\mathbbm{i}q_{\uparrow}q_{\downarrow}\cos(\phi)t(\eta\gamma(\alpha^*-\alpha)-2 w ).
\end{eqnarray}
We perform a variable transformation in order to remove
the first order derivatives in Eq.~(\ref{eq:1})
\begin{eqnarray}
 \alpha&=&z \mathbbm{e}^{s(\kappa-\mathbbm{i}
\Omega)}, \:\:\: \alpha^*=z^* \mathbbm{e}^{s(\kappa+\mathbbm{i}
\Omega)}, \:\:\: t=s,
\end{eqnarray} and obtain
\begin{eqnarray}
\partial_s{\chi}_{\sigma\sigma}&=&(\mathbbm{i}(r_{\sigma}
\gamma+K)\mathbbm{e}^{s\kappa}(z \mathbbm{e}^{-s\mathbbm{i} \Omega}+z^*
\mathbbm{e}^{s\mathbbm{i} \Omega})\nonumber\\
&-&\eta\kappa |z
|^2\mathbbm{e}^{2s\kappa})\chi_{\sigma\sigma}\nonumber\\
&-&r_\sigma\mathbbm{i}
{\delta}\chi_0(z\mathbbm{e}^{s(\kappa-\mathbbm{i}
\Omega)})F(z\mathbbm{e}^{s(\kappa-\mathbbm{i} \Omega)},s),
\end{eqnarray} 
which can be solved analytically, and transformed back to the initial variables $\alpha, t$. The solution reads
\begin{equation}
\chi_{\sigma\sigma}(\alpha,t)=\frac{|\langle\sigma|\Psi\rangle|^2}{4\pi}\chi_{\sigma\sigma}^{\rm
hom}(\alpha,t)-\frac{\mathbbm{i} r_{\sigma} \delta}{4\pi}\chi_{\sigma\sigma}^{\rm
inh}(\alpha,t),\label{chi_d}
\end{equation} 
where
\begin{eqnarray} 
&\chi_{\sigma\sigma}^{\rm hom}&(\alpha,t)=
\exp\left(-\frac{|\alpha|^2\eta}{2}
	+\mathbbm{i}(r_{\sigma}\gamma+K)
	\right.\\
&\cdot&\!\!\!\!\!\!\left.
	\left(\frac{\alpha(1-\mathbbm{e}^{-t(\kappa-\mathbbm{i} \Omega)})}{\kappa-\mathbbm{i}\Omega}
	+\frac{\alpha^*(1-\mathbbm{e}^{-t(\kappa+\mathbbm{i} \Omega)})}{\kappa+\mathbbm{i}\Omega
}\right)\right),\nonumber
\end{eqnarray} 
and
\begin{equation}
\chi_{\sigma\sigma}^{\rm inh}(\alpha,t)=\int_0^t \!\!ds\chi_{\sigma\sigma}^{\rm hom}(\alpha,s)F\left(\alpha\mathbbm{e}^{-s(\kappa-\mathbbm{i}\Omega)},t-s\right)\label{chi_diag}.
\end{equation}  

\subsubsection{The off-diagonal density matrix elements}
The method and approximations  of the previous section can be  used to solve the off-diagonal
equations. From this solution we intend to extract information about the qubit coherence $\langle \hat \sigma_x\rangle(t)$. We start with
\begin{eqnarray} 
\dot{\chi}_{\uparrow\downarrow}&=&(2\mathbbm{i}
\gamma(\partial_{\alpha^*}-\partial_\alpha)+\mathbbm{i}
\Omega(\alpha\partial_{\alpha}-\alpha^*\partial_{\alpha^*})-2\mathbbm{i} w \nonumber\\
&+&\mathbbm{i}K(\alpha+\alpha^*)+\mathcal{D})\chi_{\sigma\sigma}-\mathbbm{i} {\delta}\chi_0(\alpha)G(\alpha,t),
\end{eqnarray} where
\begin{eqnarray}
G(\alpha,t)&=&q_{\downarrow}^2-q_{\uparrow}^2-t\mathbbm{i}(\gamma-K(q_{\downarrow}^2-q_{\uparrow}^2))(\alpha+\alpha^*)\nonumber\\
&-&4t \delta q_{\uparrow}q_{\downarrow}\sin(\phi).
\end{eqnarray} 
The variable transformation in this case originates from
\begin{eqnarray}
\partial_s\alpha&=&(-\mathbbm{i}\Omega+\kappa)\alpha+2\mathbbm{i}\gamma\nonumber,\\
\partial_s\alpha^*&=&(\mathbbm{i}\Omega+\kappa)\alpha^*-2\mathbbm{i}\gamma\nonumber,
\end{eqnarray} 
and reads
\begin{eqnarray}
\alpha&=&\frac{2\mathbbm{i}\gamma}{\kappa-\mathbbm{i}\Omega}\left(\mathbbm{e}^{s(\kappa-\mathbbm{i}\Omega)}-1\right)+z\mathbbm{e}^{s(\kappa-\mathbbm{i}\Omega)}\nonumber,\\
\alpha^*&=&-\frac{2\mathbbm{i}\gamma}{\kappa-\mathbbm{i}\Omega}\left(\mathbbm{e}^{s(\kappa+\mathbbm{i}\Omega)}-1\right)+z^*\mathbbm{e}^{s(\kappa+\mathbbm{i}\Omega)},\nonumber\\
t&=&s.
\end{eqnarray} 
We obtain
\begin{eqnarray}
\partial_s{\chi}_{\uparrow\downarrow}&=&(-2\mathbbm{i} w -\eta\kappa
\alpha(z,s)\alpha^*(z^*,s)\nonumber\\
&+&\mathbbm{i}{K}(\alpha(z,s)+\alpha^*(z^*,s)))\chi_{\uparrow\downarrow}\nonumber\\
&-&\mathbbm{i} {\delta}\chi_0(\alpha(z,s))G(\alpha(z,s),s),
\end{eqnarray} 
which can be solved analytically, and transformed back
to $\alpha, t$. The solution reads
\begin{equation}
\chi_{\uparrow\downarrow}(\alpha,t)=\frac{q_{\uparrow}q_{\downarrow}\mathbbm{e}^{-\mathbbm{i}\phi}}{4\pi}\chi_{\uparrow\downarrow}^{\rm hom}(\alpha,t)
-\frac{\mathbbm{i} \delta}{4\pi}\chi_{\uparrow\downarrow}^{\rm inh}(\alpha,t),
\end{equation} 
where
\begin{eqnarray} 
&\chi_{\uparrow\downarrow}^{\rm hom}&\!\!\!\!\!(\alpha,t)=
\exp\left(-\frac{|\alpha|^2}{2}\eta
-2 \mathbbm{i}  t w -\frac{4 t \gamma  (\gamma  \eta  \kappa -\mathbbm{i} K  \Omega )}{\kappa^2+\Omega ^2}\right. \nonumber\\
&+&\!\!\!\!\!\frac{4 \gamma(\gamma\eta(\kappa^2-\Omega ^2)-2 \mathbbm{i}  K \kappa \Omega)}{(\kappa^2+\Omega ^2)^2}\label{chi_off1}\\
&+&\!\!\!\!\!\frac{K +\gamma  \eta }{\kappa+\mathbbm{i} \Omega }\left(\mathbbm{i}(1-\mathbbm{e}^{-t (\kappa +\mathbbm{i} \Omega )}) \alpha^*
-\frac{2 \mathbbm{e}^{-t (\kappa +\mathbbm{i} \Omega )} \gamma}{\kappa+\mathbbm{i}\Omega}\right)\nonumber\\
&+&\!\!\!\!\!\left.
\frac{K -\gamma  \eta}{\kappa-\mathbbm{i} \Omega}\left(\mathbbm{i} (1-\mathbbm{e}^{-t(\kappa-\mathbbm{i}\Omega)})\alpha
+\frac{2 \mathbbm{e}^{-t(\kappa-\mathbbm{i}\Omega)} \gamma }{\kappa-\mathbbm{i}\Omega}\right)\right), \nonumber
\end{eqnarray}  
and
\begin{eqnarray} 
&\chi_{\uparrow\downarrow}^{\rm inh}&\!\!\!\!\!(\alpha,t)=
\int_0^t ds \chi_{\uparrow\downarrow}^{\rm hom}(\alpha,s)\label{chi_off2}\\
&&G\left(\mathbbm{e}^{-s(\kappa-\mathbbm{i} \Omega) } \alpha +\frac{2\left(1-\mathbbm{e}^{-s(\kappa-\mathbbm{i} \Omega)}\right) \gamma }{\mathbbm{i}\kappa +\Omega },t-s\right)\nonumber.
\end{eqnarray}  

From the density matrix calculated above we can extract information about the qubit relaxation and dephasing during the short interaction with the dissipative oscillator.
\section{Conversion to circuit parameters}\label{parameters}
In the following we give a recipe \cite{Tian02} to obtain the parameters entering the calculation of this paper from the circuit components
\begin{eqnarray*}
 \Omega=\sqrt{\frac{2\pi I_c^{\rm eff}}{C_S\Phi_0}}\left(1-\left(\frac{I_B}{I_{\rm c}^{\rm eff}}\right)^2\right)^{\frac{1}{4}},
 &&
 m=\left(\frac{\Phi_0}{2\pi}\right)^2C_S,  \\
 \gamma =-\frac{M_{qS}I_{\rm q}I_B\tan\phi_m^0}{4\mu},&&
\kappa=\frac{1}{2RC_S},\\  
\tan\phi_m^0=\frac{I_B}{\sqrt{{I_{\rm c}^{\rm eff}}^2-I_B^2}},&&K=\frac{I_B}{2e}\sqrt{\frac{\hbar}{2 m \Omega}},
\end{eqnarray*}
where $\Phi_0=h/2e$ is the magnetic flux quantum for a 
superconductor, $M_{qS}$ is the qubit-SQUID  mutual inductance,  $I_{\rm c}^{\rm eff}$ is the effective critical current of the SQUID at the particular flux bias, $I_B$ is the amplitude of the dc bias pulse applied to the SQUID, $C_S$ the SQUID shunt capacitance, $R$ the internal resistance of the measurement circuitry, and $I_{\rm q}$ is the circulating current of the localized states of the qubit. The expression for $K$ is derived in the limit of a small geometric inductance, low critical current and large shunt capacitor where one can approximate the SQUID dynamics as that of a single Josephson junction with a variable critical current.  The momentum of the oscillator $p$ and the voltage across the SQUID are related by
\begin{eqnarray}
V&=&\frac{ep}{C_S\hbar},
\end{eqnarray}
where $e$ is the electron charge.
The parameters used to generate Figs. \ref{decoherence}, \ref{spiral}, \ref{detection}, \ref{detection2} are
\begin{eqnarray*}
I_{\rm c}^{\rm eff}=0.5\cdot 10^{-6} {\rm A},&& I_B=0.87 I_{\rm c}^{\rm eff},\\
C_S=2\cdot 10^{-11}{\rm F},&& M_{qS}=100\cdot 10^{-12} {\rm H},\\
I_{\rm q}=438\cdot 10^{-9}{\rm A}, &&\tau=0.3\cdot 10^{-9}{\rm s},\\
\delta/(2\pi)&=&0.8\cdot 10^{7}{\rm Hz}.
\end{eqnarray*}

 \end{document}